# Exotic compositional ordering in Mn-Ni-As intermetallics

Bruno Gonano[a], Øystein Slagtern Fjellvåg[b], Gwladys Steciuk[c], Dipankar Saha[a], Denis Pelloquin[d] and Helmer Fjellvåg[a]

[a]Center for Materials Science and Nanotechnology, Department of Chemistry, University of Oslo, P.O. Box 1033 Blindern, N-0315 Oslo, Norway

[b]Department for Neutron Materials Characterization, Institute for Energy Technology, PO Box 40, NO-2027, Kjeller, Norway.

[c]Institute of Physics, Academy of Sciences of the Czech Republic, v.v.i, Na Slovance 2, Prague 18221, Czech Republic

[d]Laboratoire CRISMAT, UMR 6508 CNRS ENSICAEN, 6 bd du Maréchal Juin, 14050 Caen Cedex 4, France

**Recent advances in tools for crystal structure analysis enabled us to describe a new phenomenon in structural chemistry, which, to this day, has remained hidden. Here we describe a crystal structure with an incommensurate compositional modulation, $Mn_{0.6}Ni_{0.4}As$. The sample adopts the NiAs type structure, but in contrast to a normal solid solution, we observe that manganese and nickel separate into layers of MnAs and NiAs with thickness of 2-4 face-shared octahedra. Experimentally, results are obtained by combination of 3D electron diffraction, scanning transmission electron microscopy and neutron diffraction. The distribution of octahedral units between the manganese and nickel layers is perfectly described by a modulation vector $q = 0.360(3)\ c^*$. An additional periodicity is thus present in the compound. Positional modulation is observed of all elements as a consequence of the occupational modulation.**

The nature of atomic arrangements has still its secrets. Since the discovery of X-rays, scientists have studied atomic arrangements in solids considering their intrinsic beauty and their role as the active link between atoms and physical properties of materials. In a crystal structure, one finds different sites for cations and anions, reflecting their different chemical properties (*i.e.* size, charge, electronegativity, *etc.*). This gives rise to a huge range of crystal structure types, from simple to very complex ones. MnAs (space group *P6₃/mmc*) is an example of a simple crystal structure derived from ABAB sphere packing of As, with Mn in octahedral sites formed by closed-packed As-layers, giving rise to face-sharing chains of $MnAs_6$-octahedra along [001] with two octahedra per unit cell. In the *ab*-plane, $MnAs_6$-octahedra are edge-sharing. NiAs (space group *P6₃/mmc*) adopts the same structure type, but with different unit cell parameters (due to the different size of Mn and Ni).

The intermediate compositions between MnAs and NiAs, $Mn_{1-x}Ni_xAs$ ($0 \leq x \leq 1$), were investigated in the 1980s and revealed an unresolved structural phenomenon for $0.25 \leq x \leq 0.75$[1]. A possible modulation due to Ni ordering was proposed, but no available technique could confirm or invalidate this speculation. Therefore, the mystery of Ni-Mn-ordering in $Mn_{1-x}Ni_xAs$ remained unresolved and forgotten. Today, state-of-the-art analysis tools have evolved to the point that revisiting this system is likely to provide answers.

The study of the lattice deduced from conventional electron diffraction (ED) data revealed an expected complex unit cell. Electron diffraction pattern of the [100]-zone axis of $Mn_{0.6}Ni_{0.4}As$ shows additional reflections (**Fig. 1a**) that can be indexed by an incommensurate modulation vector ($q = 0.36\ c^*$), confirming the presence of additional order in the compound. The corresponding High-Angle Annular Dark Field (HAADF) image coupled with local EDX collections yields good elemental contrasts correlated to Mn and Ni species and reveals two types of layers perpendicular to the modulation vector [001] (**Fig. 1b and c**). We interpret them as a specific ordering between Mn and Ni based layers, as supported by energy-dispersive X-ray spectroscopy mapping (EDX) (**Fig. 1b**). The thickness of these layers can be measured by the number of connected octahedra, being 3-4 for the MnAs and 2-3 for the NiAs part of the integrated structure.

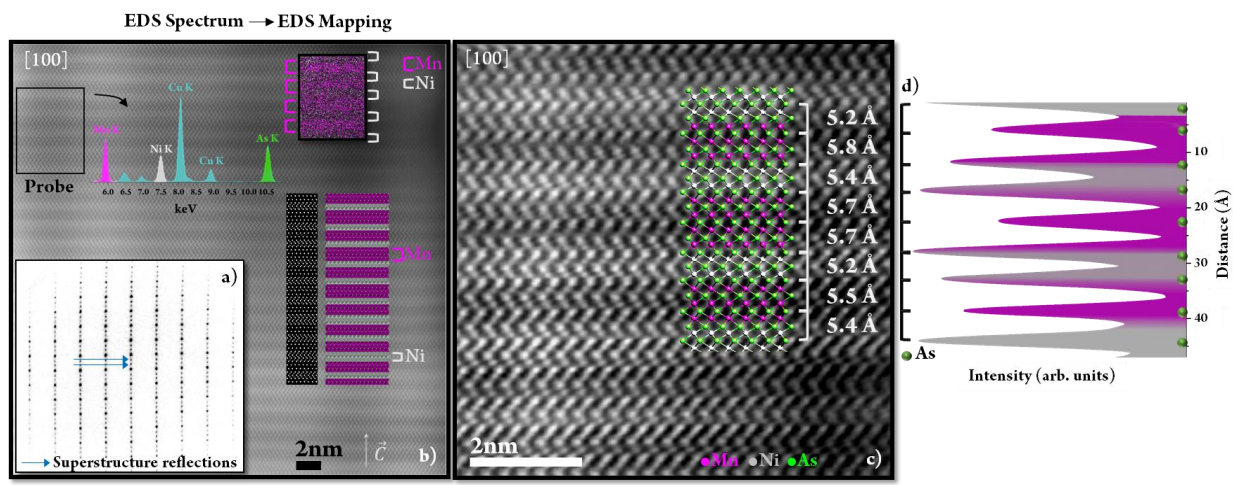

**Figure 1:** a) Electron diffraction of the [100]-zone axis of $Mn_{0.6}Ni_{0.4}As$ showing satellite reflections, corresponding to an incommensurate modulation vector along $c^*$ ($q = 0.36\ c^*$); b) Experimental [100] HAADF image of $Mn_{0.60}Ni_{0.4}As$ (magnification x8M). Bright dots are related to As rows while darker and brighter layers represent MnAs and NiAs, respectively. On the right side, simulated image confirms the goodness of the structural model. Latter is inserted beside. Top left insert: EDS spectrum related to recorded zone (Probe) identified by a black box. Top right insert: EDS mapping evidencing richer zones in Mn (purple) and Ni (grey); c) Experimental [100] HAADF image of $Mn_{0.60}Ni_{0.4}As$ (magnification x25M) showing variations in distances between As atoms from Mn (darker) and Ni (brighter) layers. A snapshot of the structural model is inserted; d) Intensity line profile extracted from the HAADF image along [001], displaying the variations of As-As distances in the MnAs and NiAs layers.

This type of intrinsic "nano-layering" is unique. It is reminiscent of artificially created ultra-thin heterostructures; it may have features in common with spinoidals; yet it is different and unique with the compositional modulation being a genuine part of the crystal structure of the phase. It is also unlike those of other modulated intermetallics, where typically two different networks are present in the same compound, creating two sets of periodicities, which thus require higher-dimensional formalism to be described[2,3]. Note, the latter modulations are not occupational in nature.

Careful analysis of the As-As distances along [001] in an x25M HAADF image, reveals a distinct difference between the zones richer in Mn or in Ni (**Fig. 1c**). Whereas for the brighter layer (Ni-rich) the As-As distance tends to be shorter, it increases in the darker Mn-rich regions, according to the longer $c$-axis of MnAs (5.8 Å) compared to NiAs (5.0 Å). This is supported by the extracted line profile for the As-As distances in **Fig. 1d**.

At this point, it is clear that the Mn-Ni ordering is real. We now use 3D ED to unveil a structural model.[4] This technique has recently proved his ability to yield valid structural solutions in complex systems[5] and provide single-crystal diffraction data on small areas of few hundreds of nanometers, also on powder samples . It has to be mentioned here that 3D ED represents a broad range of experimental protocols and refer in this study to Precession Electron Diffraction Tomography (PEDT). 3D ED shows the expected hexagonal subcell, but also additional reflections at incommensurate positions (**Fig. 2a**). The reciprocal space was indexed using the superspace formalism considering a hexagonal unit cell: $a = 3.653(8)$ Å, $c = 5.417(3)$ Å and with a modulation vector $q = 0.360(3)\ c^*$ to index the satellite reflections up to first order. As expected from the conventional ED, the modulation is incommensurate. The volume of the average unit cell ($V = 62.67$ Å$^3$) is in between the volumes of NiAs ($V = 56.7$ Å$^3$) and MnAs ($V = 68.55$ Å$^3$).[6,7]

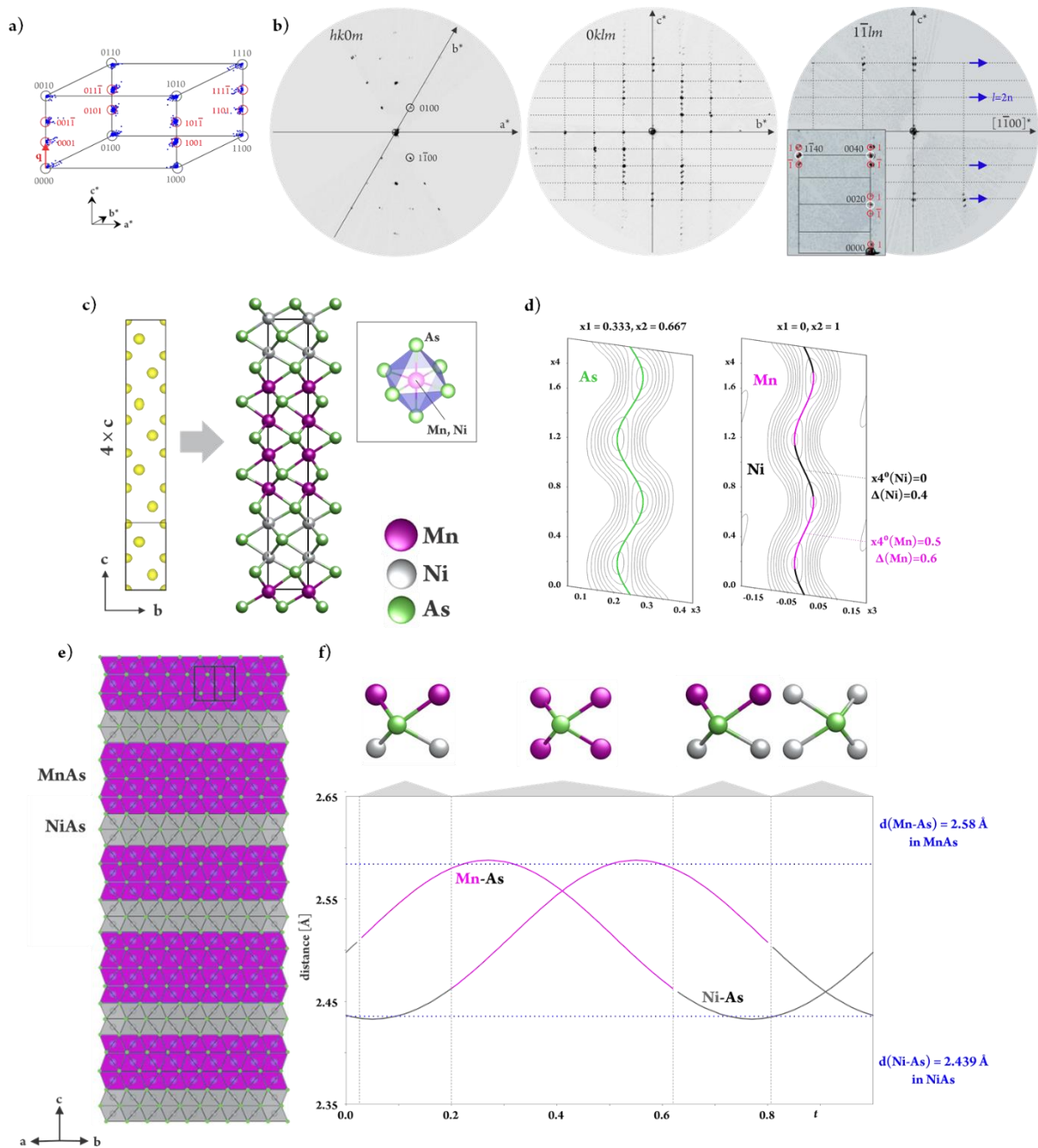

**Figure 2: a)** Reciprocal space projected in one-unit cell showing satellite reflections up to first order (red); **b)** Sections of reciprocal space from PEDT data; **c)** [100] projection of the 3D electronic potential map and its interpretation as a structural model. For better visualization of the layers, the model is extended along the stacking direction [001]. The Mn and Ni cations are in octahedral coordination; **d)** On the De Wolf section $x3$-$x4$ calculated around the As site, the cationic site is described with a continuous harmonic function and split between 2 sites for Mn and Ni using discontinuous crenel-like functions; **e)** Extended [110] projection of the refined structure against PEDT data (dynamical refinement); **f)** Variation in Mn – As and Ni - As distances showing the evolution of the local As environment as function of the modulation.

Information on symmetry is obtained from *e.g.* the $(1-1lm)^*$ and $(0klm)^*$ sections of the reciprocal space for which the extinction conditions $l = 2n$ on $(00l)$ and $(1-1lm)$ are characteristic of a $6_3$-screw axis along [001] and a *c*-glide mirror perpendicular to [1-10], respectively (**Fig. 2b**). These conditions are compatible with two superspace groups (SSG); $P6_3/mmc(00\gamma)0000$ and $P6_3mc(00\gamma)000$. Our subsequent refinements indicated that the structure is centrosymmetric $P6_3/mmc(00\gamma)0000$. The structure was finally solved from the 3D ED data in SSG $P6_3/mmc(00\gamma)0000$ with ***q*** = 0.360(3) ***c*** and a data coverage of 100% for 0.7 Å$^{-1}$ resolution. The main experimental parameters are listed in **Table 1**.

The initial solution obtained by charge flipping is a 3D map of the electrostatic potential (e-map) represented as isosurfaces (**Fig. 2c**). For better visualization of the Mn-Ni stacking sequence, the e-map, as well as the interpreted model, is represented in an averaged supercell $a \times b \times 4c$. This is further elaborated by calculating De Wolf sections *x3-x4* around the two atomic sites corresponding to Mn/Ni and As atoms (**Fig. 2d**). First, the average framework of (Mn, Ni)As$_6$ octahedra is revealed in agreement with the initially reported structure.[1] However, it is evident that the As site exhibits a displacive modulation along *c*, described using a continuous harmonic function. In order to account for the Mn/Ni ordering along [001], the site shared by Mn and Ni atoms is split using discontinuous crenel-like functions associated with one harmonic function.[8]

Within the resolution of the 3D ED data, the electron scattering amplitudes for Mn and Ni are too close to reliably detect a difference in the electrostatic potential between the domains. For this reason, the centres of the crenels are set to $x4^0$(Mn) = 0.5 and $x4^0$(Ni) = 0 in order to strain the shortest (longest) distances for Ni-As (Mn-As). The crenel widths were set according to the composition: $\Delta$(Mn) = 0.6 and $\Delta$(Ni) = 0.4 (**Fig. 2d**). The interpreted model exhibits the alternation of two types of blocks with variable thickness consisting either of layers of MnAs$_6$ octahedra or of layers of NiAs$_6$ octahedra.

Dynamical refinements were carried out against the 3D ED data.[10,11] The included reflections were chosen following the recommendation by Palatinus *et al*.[10] $RS$g(max) = 0.9, S(max)g(matrix) = 0.01 Å$^{-1}$, g(max) = 1.7 Å$^{-1}$. Because of the rather small unit cell volume and the high symmetry, the $RS$g(max) was set to a high value to thereby include a sufficiently large number of reflections and maintain a good reflection-to-parameter ratio (**Table 1**). The refinement gave reliability factors of *R/wR*(obs/all) = 0.1385/0.1465 for *N* obs/all = 796/1788 [main: *R/wR*(obs) = 0.1267/0.1428; 1. order: *R/wR*(obs) = 0.1624/0.1574]. This represent a huge improvement compared to what offered in a kinematical refinement [*R/wR*(obs/all) = 0.03018/0.03539 for *N* obs/all = 88/128].

The amazing layering of Mn$_{0.6}$Ni$_{0.4}$As with alternation of 3-4 layers of MnAs and 2-3 layers of NiAs (**Fig. 2e**), results in a genuine incommensurate modulation with respect to the chemical composition. The simulated HAADF image (calculated by JEMS software) based on the aforementioned incommensurate structure model is included in **Fig. 1b** and shows very good compliance with the observed image. The observed positional modulation of the Mn, Ni and As atoms along [001] dictates a difference in the cell volume for the MnAs and NiAs layers, being in agreement with unit cell volumes of the binary compounds as well as the HAADF images. The variation of Mn-As and Ni-As bond lengths caused by the positional modulation of As and Mn/Ni is illustrated in **Fig. 2f**. This information corroborates our initial observations from HAADF imaging (**Fig. 1b and 1c**).

The collected powder neutron diffraction (PND) pattern shows intense reflections at low scattering angle (**Fig. 3**), *e.g.* at 0.42 Å$^{-1}$, which is a first-order satellite reflection from the compositional modulation of Mn and Ni. The strong signal of the satellite reflection is a consequence of the excellent contrast between Mn and Ni in PND, due to the different signs of their scattering lengths. Furthermore, we refined the crenel widths during Rietveld refinements and achieved a high accuracy for the sample composition, unlike what was feasible with 3D ED data. The derived composition of Mn$_{0.598(3)}$Ni$_{0.402(7)}$As matches perfectly the nominal composition from synthesis. The modulation vector obtained from the PND Rietveld refinements [*q* = 0.3594(2) *c**] is identical to that obtained by 3D ED [*q* = 0.360(3) *c**]. We note that the results obtained from refinements of PND and 3D ED data are identical, within statistical uncertainty. We emphasize that PND is superior for refining the exact composition, while 3D ED is more powerful for structural determination. The final refined structural model from both data sets is presented in **Table 2**.

**Table 1.** Summary of data collection conditions and refinement parameters

| Structural formula | AsMn$_{0.6}$Ni$_{0.4}$ |
|---|---|
| Unit–cell parameters (PEDT) | $a$ = 3.653(8) Å, $c$ = 5.417(3) Å, $\gamma$ = 120°, $q$ = 0.360(3) $c^*$, $V$ = 62.67 Å$^3$ |
| Unit–cell parameters (NPD) | $a$ = 3.6520(2) Å, $c$ = 5.4280(6) Å, $\gamma$ = 120°, $q$ = 0.3594(2) $c^*$, $V$ = 62.694(6) Å$^3$ |
| Z | 2 |
| Density [g.cm$^{-3}$] (from NPD) | 6.96(7) |
| Space group | $P6_3/mmc$ |
| Temperature | Ambient T |
| Diffractometer | JEOL2010 |
| Radiation (wavelength) | electrons, (0.0251 Å) |
| Resolution | 0.1–0.7 Å$^{-1}$ |
| Limiting Miller indices | –2<h<0, 0<k<4, 0<l<7, -1<m<1 |
| No. of independent reflections (obs/all) – kinematic | all : 86/128<br>main : 38/45<br>order 1 : 48/83 |
| $R_{int}$ (obs/all) – kinematic | 0.5143/0.5212 |
| Redundancy | 6.641 |
| No. of collected reflections (obs/all) – dynamical | all : 1666/6635 |
| Coverage for sin$\theta/\lambda$ = 0.7Å$^{-1}$ | 100 % |
| Dynamical refinement | |
| RSg(max) | 0.9 |
| No. of reflections (obs/all) | all: 796/1788<br>main: 437/622<br>order 1: 359/1166 |
| $R, wR$ (obs) | all: 0.1385/0.1465<br>main: 0.1267/0.1428<br>order 1: 0.1624/0.1574 |
| Crystal thickness | 300(3) Å |
| N parameters/N struct. parameters | 76/6 |
| Rietveld refinement (Neutron Powder data) | |
| No. of independent reflections (obs/all) | all: 41/44<br>main: 9/9<br>order 1: 18/18<br>order 2: 14/17 |
| $R, wR$ (obs) | all: 0.0571/0.0659<br>main: 0.0574/0.0571<br>order 1: 0.0401/0.0501<br>order 2: 0.1229/0.1171 |
| N struct. parameters | 5 |
| Profile: | Rp = 0.0654, wRp = 0.0894, GOF = 0.0334 |

**Table 2.** Positional parameters

| atom | Δ/Occ. | harm. | x/a | y/b | z/c | Uiso [Å$^2$] |
|---|---|---|---|---|---|---|
| \multicolumn{7}{|c|}{Dynamical refinement against 3D ED data} |
| Mn1 | 0.6 | | 0 | 1 | 0 | 0.0191(9) |
| | | s,1 | 0 | 0 | -0.0313(6) | |
| Ni1 | 0.4 | | 0 | 1 | 0 | 0.0191(9) |
| | | s,1 | 0 | 0 | -0.0313(6) | |
| As1 | 1 | | 0.3333 | 0.6667 | 0.25 | 0.0215(9) |
| | | s,1 | 0 | 0 | -0.0468(6) | |
| \multicolumn{7}{|c|}{Refined positional parameters (powder neutron)} |
| Mn1 | 0.598(3) | s,1 | 0 | 0 | -0.003(3) | 0.016(1) |
| Ni1 | 0.402(3) | s,1 | 0 | 0 | -0.003(3) | 0.016(1) |
| As1 | 1 | s,1 | 0 | 0 | -0.037(1) | 0.005 |

ADP harmonic parameters (3D ED)

| atom | U11 | U22 | U33 | U12 | U13 | U23 |
|---|---|---|---|---|---|---|
| Mn1 | 0.0204(12) | 0.0204(12) | 0.0164(12) | 0.0102(6) | 0 | 0 |
| Ni1 | 0.0204(12) | 0.0204(12) | 0.0164(12) | 0.0102(6) | 0 | 0 |
| As1 | 0.0179(11) | 0.0179(11) | 0.0288(14) | 0.0089(6) | 0 | 0 |

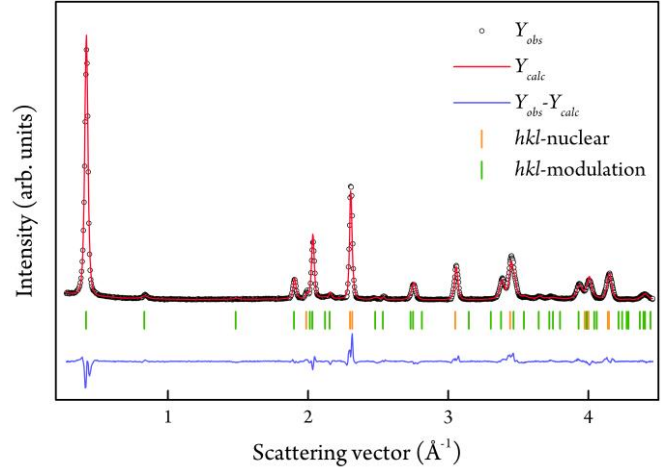

**Table 1: Parameters for data collection and refinement of 3D ED and neutron diffraction data. Table 2. Positional parameters extracted from the dynamical refinement of 3D ED data. Figure 3: Experimental (black crosses), calculated (red line) and difference (blue line) neutron powder diffraction pattern of Mn$_{0.6}$Ni$_{0.4}$As according to Rietveld refinement. Main and satellite Bragg peak positions are indicated with green and orange sticks, respectively.**

We have here described for the first time an exotic phenomenon of composition modulation intimately coupled to a structural modulation. The Mn-Ni composition modulation in the solid solution compound, Mn$_{0.6}$Ni$_{0.4}$As, ideally having a simple basic NiAs-type structure, represents to the best of our knowledge a unique case. We stress that this modulation of composition, materialized by ultrathin layers consisting of edge- and face-sharing Mn-As octahedra and of Ni-As octahedra, is different from any stacking or order/disorder phenomena. The modulated structure is the genuine crystal structure of Mn$_{0.6}$Ni$_{0.4}$As. The well-defined periodic variation in the stacking of MnAs and NiAs 2D-layers creates an incommensurate modulation, with 3-4 thick layers of MnAs and 2-3 layers of NiAs. We showed that the structural model to describe this peculiar phenomenon can be derived from 3D ED and validated by refinement of powder neutron diffraction data, with a derived composition in full compliance with nominal sample composition from synthesis. This model, as well as the coupled HAADF-EDS mapping images, shows that the MnAs layers have a longer $c$-axis than the NiAs layers, in full agreement with data for the individual binary compounds.

We have demonstrated the existence of a hitherto unknown nanophenomenon a in a solid-solution phase, and the ability, strength and robustness of state-of-the-art tools in crystallography and methodology to solve and describe complex incommensurate phenomena, all acting as inspiration for future research.

## Acknowledgments

This work is part of the activities of the NAMM project (Novel Approaches to Magneto-Structural phases transitions in Metallic systems), supported by the Research Council of Norway (Grant no. 263241). Authors want to acknowledge Susmit Kumar, Center for Materials Science and Nanotechnology, Department of Chemistry, University of Oslo for fruitful discussions. The authors acknowledge Vivian Nassif, Inès Puente-Orench and the staff at the D1B beamline, ILL, France, for beamtime allocation and technical support. (S)TEM experiments have been performed in the CRISMAT Lab. (UMR6508, Caen, France) within the frame of the METSA federation (FR3507). G.S. acknowledges the Czech Science Foundation through Project No. 19-07931Y.

## Conflict of Interest

The authors declare no conflict of interest.

## Methods

### Synthesis

The $Mn_{0.6}Ni_{0.4}As$ compound was synthesized using solid-state reaction. First, MnAs and NiAs binaries were synthesized from stoichiometric amounts of elements Mn, Ni and As, weighed and crushed using agate mortar and pestle. Powders were introduced in alumina crucibles and put into vacuum sealed quartz tubes. Latter were placed in a standing tube furnace and slowly heated up to 900°C (0.1°C/min) for a week. Cooling down to room temperature was performed at a 1°C/min rate. Ternary compound was prepared weighing stoichiometric amounts of MnAs and NiAs introduced in alumina crucibles, latter placed in quartz tubes, introduced in a standing furnace. Tube was slowly heated (1°C/mn) up to 800°C, held at this temperature for 5 days and cooled to room temperature at the same rate. This process has been repeated three times with intermediate crushings. Finally, the sample was annealed at 600°C for 1 month with the same heating and cooling rate.

### Atomic imaging

High-Angle Annular Dark Field (HAADF) imagery imaging was performed on an ARM 200F with a corrected probe, operating at 200kV and equipped with Centurio EDX spectrometer. The simulated ADF images have been calculated with the JEMS software considering the convolution of the STEM probe with the intensity of the object (square of the projected potential multiplied by the electron-matter interaction constant of the structure).

### 3D Electron Diffraction (3D ED)

A small quantity of $Mn_{0.6}Ni_{0.4}As$ powder (<1 µg) was dispersed in a butanol solution and ground in an agate mortar. A drop of the suspension was deposited and dried on a copper grid with a thin film of holey amorphous carbon. 3D ED experiment was performed with a JEOL 2010 transmission electron microscope (operating at 200 kV with a $LaB_6$ cathode) equipped with a Nanomegas DigiStar precession module and a retractile side-entry Gatan ORIUS 200D CCD camera. PEDT data of non-oriented patterns were collected at room temperature. The precession angle was set to 1.2° with a goniometer tilt step below 1°. The PEDT data set was analyzed using the computer programs PETS2.0[12], SUPERFLIP[13], and JANA2006.[14] Details about the methodology to solve incommensurately modulated structures using PEDT can be found elsewhere[15–18]. For each data set, the result is a list of *hklm* indices with associated intensities and estimated standard deviations based on counting statistics (**Table 1**).

### Neutron Diffraction

Neutron powder diffraction (NPD) data has been collected at room temperature on D1B, ILL, France, using wavelength λ = 2.52 Å. About 5 g of the sample was put into a cylindrical vanadium can. Rietveld refinements were carried out in JANA2006.[14] The Rietveld refinement resulted in *Rp* = 0.0654, *wRp* = 0.0894, *GOF* = 0.0334 and *R/wR* (obs) all = 0.0571/0.0659 for *N* obs/all reflections = 41/44 (main: *R/wR*(obs) 0.0574/0.0571; order 1: *R/wR*(obs) = 0.0401/0.0501; order 2: *R/wR*(obs) = 0.1229/0.1171) (see refinement details in **Table 1**). Data are found at http://doi.ill.fr/10.5291/ILL-DATA.DIR-172